\documentclass[aps,prx,twocolumn,superscriptaddress,amsmath,amssymb,10pt,epsfig,nofootinbib]{revtex4-2}

\usepackage{appendix}
\usepackage{amssymb, amsmath,amsthm,bm}    
\usepackage{graphicx}   
\usepackage{verbatim}   
\usepackage{color}      
\usepackage{subcaption} 
\usepackage{bbm}        
\usepackage[colorlinks=true]{hyperref}
\hypersetup{
    bookmarks=true,         
    unicode=false,          
    pdftoolbar=true,        
    pdfmenubar=true,        
    pdffitwindow=false,     
    pdfstartview={FitH},    
    pdfsubject={},   
    pdfcreator={},   
    pdfproducer={}, 
    pdfkeywords={} {} {}, 
    pdfnewwindow=true,      
    colorlinks=true,       
    linkcolor=magenta, 
    citecolor=blue,        
    filecolor=magenta,      
    urlcolor=blue           
}
\usepackage{caption}
\usepackage{tikz}
\usepackage{pgflibraryarrows}
\usepackage{pgflibrarysnakes}
\usepackage{lipsum}
\usepackage{mathtools}
\usepackage{dirtytalk}
\usepackage{ae} 
\usepackage[T1]{fontenc}
\usepackage{epsfig}
\usepackage{enumerate}
\usepackage{subcaption}
\usepackage{varioref}
\usepackage{cleveref}
\usepackage{makeidx}
\usepackage[english]{babel}
\usepackage[normalem]{ulem}
\usepackage{physics}
\usepackage{bbold}
\usepackage{float}
\usepackage{footnote}
\usepackage{gensymb}

\usepackage{algorithm}
\usepackage{algpseudocode}
\algblock{Input}{EndInput}
\algnotext{EndInput}
\algblock{Output}{EndOutput}
\algnotext{EndOutput}

\newcommand{\beq}{\begin{equation}}
\newcommand{\eeq}{\end{equation}}
\newcommand{\beqq}{\begin{equation*}}
\newcommand{\eeqq}{\end{equation*}}
\newcommand\bea{\begin{array}}
\newcommand\eea{\end{array}}
\newcommand\beaa{\begin{array*}}
\newcommand\eeaa{\end{array*}}
\newcommand\beal{\begin{align}}
\newcommand\eeal{\end{align}}
\newcommand\beall{\begin{align*}}
\newcommand\eeall{\end{align*}}

\def\o{{\omega}}

\def\L{\Lambda}

\def\s{\sigma}

\def\g{\gamma}

\newcommand{\eps}{\varepsilon}

\newcommand{\f}{\frac}

\DeclareMathOperator{\sign}{sgn}

\def\[{\left[}
\def\]{\right]}
\def\({\left(}
\def\){\right)}
\def\<{\langle}
\def\>{\rangle}

\definecolor{darkblue}{cmyk}{0.9,0.9,0,0}
\definecolor{greennote}{RGB}{0,135,41}

\begin{document}
\begin{flushright}
\href{https://arxiv.org/abs/2412.15330}{arXiv:2412.15330}
\end{flushright}

\title{Thermopower across Fermi-volume-changing quantum phase transitions\\without translational symmetry breaking}

\author{Peter Lunts}
\email{plunts@fas.harvard.edu}
\affiliation{Department of Physics, Harvard University, Cambridge MA 02138, USA}
\author{Aavishkar A. Patel}
\affiliation{Center for Computational Quantum Physics, Flatiron Institute, 162 5th Avenue, New York, NY 10010, USA}
\author{Subir Sachdev}
\affiliation{Department of Physics, Harvard University, Cambridge MA 02138, USA}
\date{\today}

\begin{abstract}
We describe the evolution of low-temperature thermopower across Fermi-volume-changing quantum phase transitions in Kondo lattice models without translational symmetry breaking. This transition moves from a heavy Fermi liquid with a conventional Luttinger-volume large Fermi surface to a ‘FL*’ state, characterized by a small Fermi surface and a spin liquid with fractionalized excitations. The onset of the large Fermi surface phase is driven by the condensation of a Higgs field that carries a unit gauge charge under an emergent U(1) gauge field. We consider the case with spatially random Kondo exchange, as this leads to strange metal behavior in electrical transport. We find a large asymmetric thermopower in a `skewed' marginal Fermi liquid, with similarities to the skewed non-Fermi liquid of Georges and Mravlje (\href{https://doi.org/10.1103/PhysRevResearch.3.043132}{Phys. Rev. Research {\bf 3}, 043132 (2021)}). Our findings are consistent with recent observations in heavy fermion compounds (Z.-Y. Cao {\it et al.\/}, \href{https://arxiv.org/abs/2408.13604}{arXiv:2408.13604}), and describe an enhancement of thermopower on the large Fermi surface side as well as a non-monotonic behavior on the small Fermi surface side.

Our results also apply to single-band Hubbard models and the pseudogap transition in the cuprates. In the ancilla framework, single-band models exhibit an {\it inverted\/} Kondo lattice transition: the small Fermi surface pseudogap state corresponds to the condensed Higgs state. This inversion results in an enhancement of thermopower on the pseudogap side in our theory, consistent with observations in the cuprates (C. Collignon {\it et al.\/}, \href{https://journals.aps.org/prb/abstract/10.1103/PhysRevB.103.155102}{Phys. Rev. B {\bf 103}, 155102 (2021)}; 
A. Gourgout {\it et al.\/}, \href{https://journals.aps.org/prx/abstract/10.1103/PhysRevX.12.011037}{Phys. Rev. X {\bf 12}, 011037 (2022)}). We argue that these observations support a non-symmetry-breaking Fermi-volume-changing transition as the underlying description of the onset of the pseudogap in the cuprates. 
\end{abstract}

\maketitle

\section{Introduction}

A number of recent experiments have explored the thermopower of metallic correlated electron compounds across Fermi-volume-changing transitions \cite{Steglich10,Nakatsuji12,Collignon21,Taillefer_Seebeck_PRX_2022,Kanoda23,Park24}. This paper aims to provide theoretical insight into these observations using a model of non-zero temperature quantum criticality across Fermi-volume-changing quantum phase transitions without translational symmetry breaking, introduced in Refs.~\cite{SVS03,SVS04} (see also Refs.~\cite{Coleman_Andrei,BGG02}). We will show that thermopower is an especially sensitive probe of key features of such non-symmetry-breaking transitions, and the enhancement and asymmetry of the thermopower observations support such a non-symmetry-breaking description of the heavy fermion compounds and of the cuprates.

There is a simple reason for thermopower to be a distinguishing diagnostic between symmetry-breaking and non-symmetry-breaking quantum phase transitions, which we mention at the outset, and describe in more detail below. In both types of transitions, electrons acquire a singular self-energy from their coupling to a critical bosonic field. The electronic self energy can have a singular particle-hole asymmetry (crucial for large thermopower) only if the boson propagator is itself particle-hole asymmetric.
The propagator in the Hertz theory for symmetry-breaking transition is particle-hole symmetric (the $-i \omega$ is absent in (\ref{eq:boson Greens function})), while non-symmetry-breaking transitions have a particle-hole asymmetry arising from the fact that the boson carries electrical charge. The system is especially sensitive to the particle-hole asymmetry in the bosonic sector because the bosons are critical, whereas any intrinsic asymmetry in the fermionic sector is small and unimportant.

Although symmetry-breaking phases may be present nearby, we will assume that the main physics in the relevant materials at intermediate temperatures is that of a Fermi volume change, and that translational symmetry breaking is a secondary, lower temperature phenomenon.
As the Luttinger relation constraints the Fermi volume by the total electron density, it follows that, in the absence of translational symmetry breaking, at least one of the metallic phases does not obey the Luttinger relation. Such non-Luttinger-volume metallic phases are permitted in the presence of fractionalization \cite{SVS03,SVS04,APAV04,Bonderson16,SS_QPM} by an extension of Oshikawa's argument for the Luttinger volume in a Fermi liquid \cite{MO00}.

It is conventional to refer to the Luttinger volume Fermi surface as a `large' Fermi surface; the phase with a large Fermi surface is a conventional Fermi liquid, and we will denote it as FL. The non-Luttinger volume Fermi surface is `small', and the phase with a small Fermi surface is a `fractionalized Fermi liquid' \cite{SVS03}, and we will denote it as FL*. We assume that fractionalization is present in the intermediate temperature FL* phase, but any symmetry breaking phases that may be present nearby at low temperatures are conventional, and do not have fractionalization.

We will describe the thermopower in the non-zero temperature quantum-critical region of the Fermi-volume-changing 
transition \cite{Kotliar01,Kotliar09,Pepin10,Shastry12,Georges_Skewed}. Kim and P\'epin \cite{Pepin10} argued that the thermopower is mostly symmetrical on the two sides of the quantum phase transition for symmetry-breaking transitions. But, as Georges and Mravlje have argued \cite{Georges_Skewed}, a large asymmetric thermopower (inside of the critical fan) can appear in other cases \cite{Kotliar01,Kotliar09,Pepin10,Shastry12,Georges_Skewed}. Our primary focus in this paper is the nature of this asymmetry, and the distinct behaviors of the non-symmetry-breaking Fermi-volume-changing transitions in Kondo lattice and single band models summarized in Figs.~\ref{fig:kondo} and \ref{fig:ancilla}, and discussed in Section~\ref{kl_vs_single} below.
\begin{figure*}
\begin{center}
\includegraphics[width=5.2in]{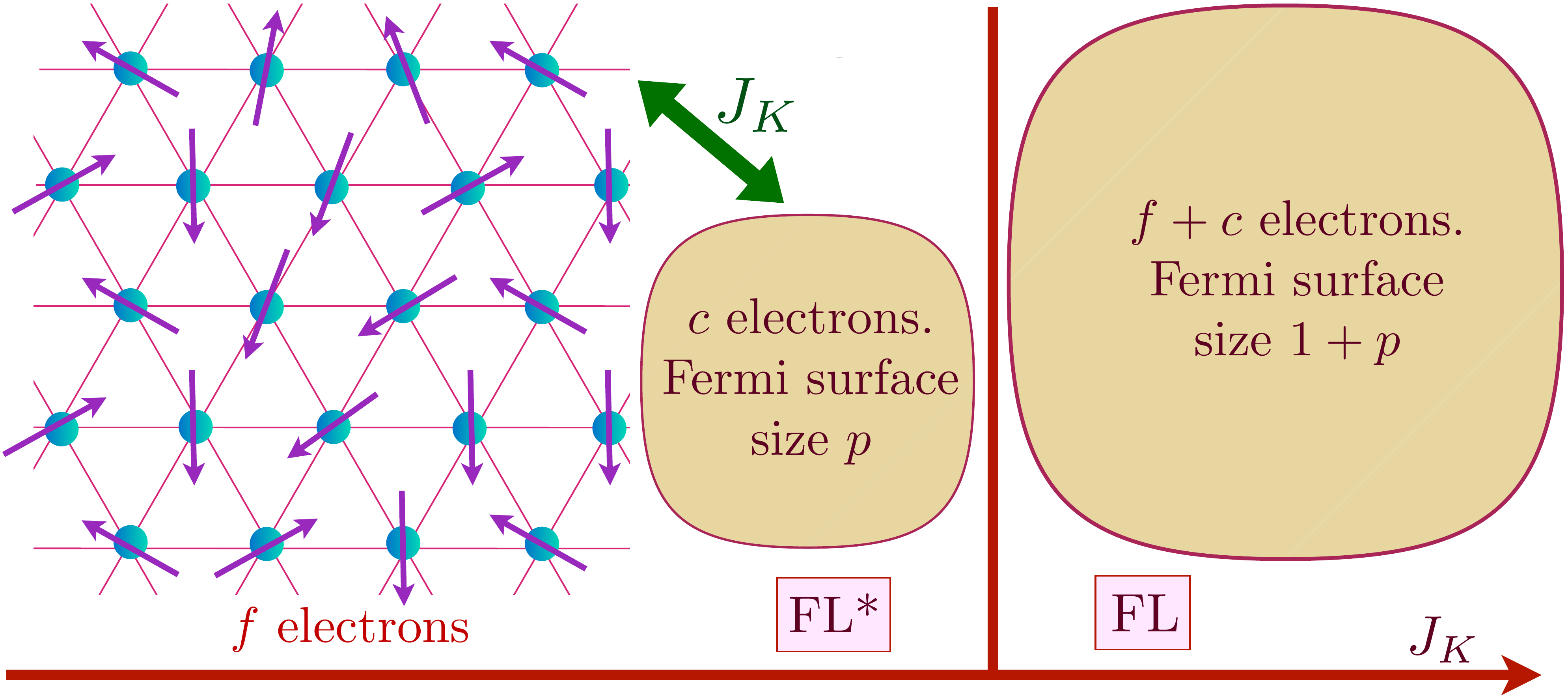}
\end{center}
\caption{Fermi volume changing transition in a Kondo lattice of spins of $f$ electrons and a conduction band of $c$ electrons (from Ref.~\cite{SSORE}). The $x$-axis is the Kondo coupling between the $c$ and $f$ electrons, denoted as $J_K$. In the fractionalized Fermi liquid (FL*) phase the $f$ electrons form a spin liquid with fractionalized spinon excitations, while the $c$ electrons form a `small' Fermi surface. In the FL phase, the $f$ and $c$ electrons hybridize, and realize a `large' Fermi surface which has the Luttinger volume of free electrons.}
\label{fig:kondo}
\end{figure*}
\begin{figure*}
\begin{center}
\includegraphics[width=6.1in]{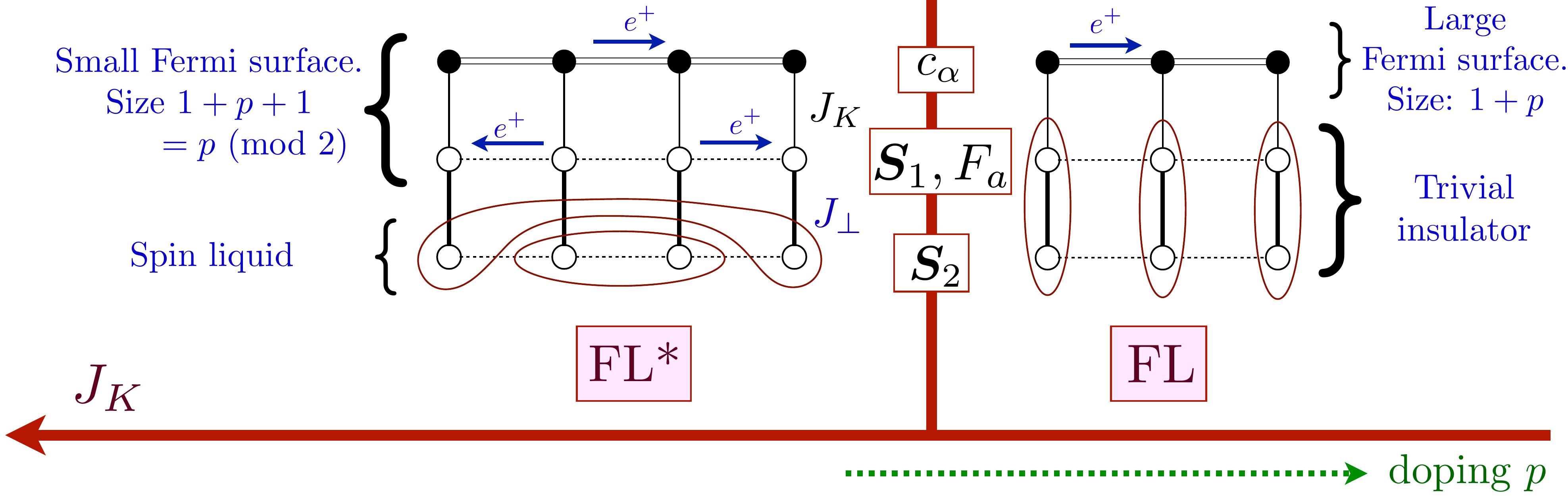}
\end{center}
\caption{Fermi volume changing transition in the ancilla theory of the single band Hubbard model (from Ref.~\cite{SSORE}; see Appendix~\ref{app:ancilla}). The physical electrons ($c_\alpha$) are in the single-band top layer, and two layers of ancilla qubits are realized by a bilayer antiferromagnet (with interlayer exchange $J_\perp$) with spins ${\bm S}_1$ and ${\bm S}_2$. The ${\bm S}_1$ are coupled to the physical electrons by the Kondo coupling $J_K$.
For $J_\perp \gg J_K$, the ancilla ${\bm S}_{1,2}$ spins form rung singlets, and a FL phase is present in the $c_\alpha$ layer. For $J_K \gg J_\perp$, we obtain the FL* phase in which the $c_\alpha$ electrons hybridize with the $F_a$ fermions representing the ${\bm S}_1$ spins to form a Fermi surface similar to the large Fermi surface of the Kondo lattice model in Fig.~\ref{fig:kondo}, while the ${\bm S}_2$ spins form a spin liquid with neutral spinon excitations. 
}
\label{fig:ancilla}
\end{figure*}
\begin{figure*}[t]
     \centering
     \includegraphics[width=5in]{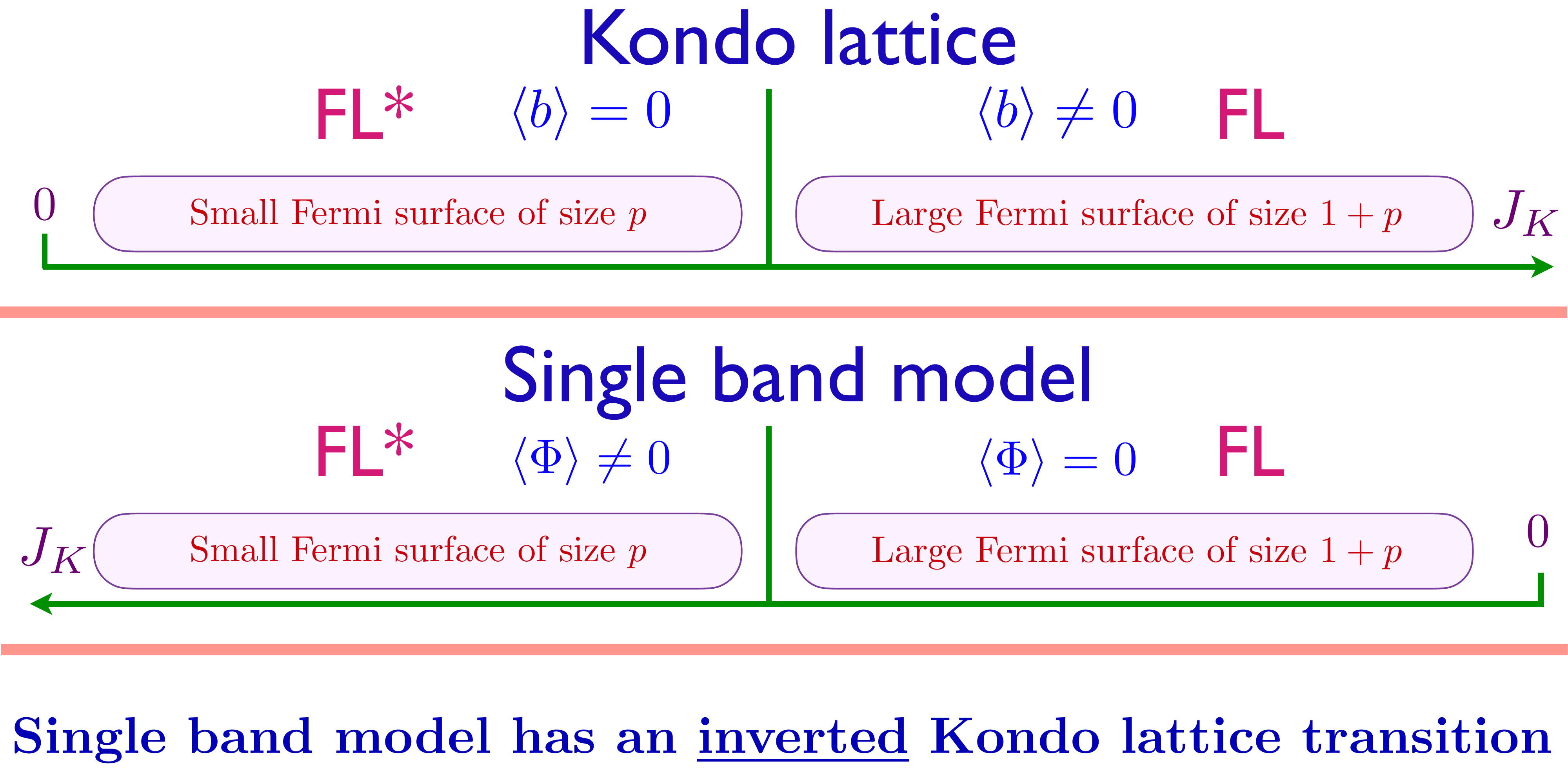}
     \caption{Contrasting the FL-FL* transition between the Kondo lattice and the ancilla theory of single-band models. For the Kondo lattice, the FL has the Higgs boson $b$ condensed. On the other hand, for the single-band model, the FL* phase has the Higgs boson $\Phi$ condensed (see Appendix~\ref{app:ancilla}).}
     \label{fig:ancilla_2}
\end{figure*}
Our specific computations are carried out in a Kondo lattice model with a spatially random Kondo coupling introduced by Aldape {\it et al.} \cite{Aldape2022}, which we present in Section~\ref{sec:theory}. It has been argued \cite{Patel2023universal,Patel2023localization,Li:2024kxr,Hardy24,PatelLuntsAlbergo2024} that including spatial randomness in the interactions is essential near quantum-critical points in metals. Spatially random interactions lead directly to spatial variations in the local position of the quantum-critical point, and such `Harris disorder' is a strongly-relevant perturbation. The quantum-critical transport properties of metals cannot be compared to observations in the absence of randomness \cite{Hartnoll:2007ih,Maslov11,Hartnoll:2014gba,Patel:2014jfa,Maslov17,Guo2022,Shi23,Haoyu_Cyclotron}. Moreover, spatially random interactions lead naturally to the observed marginal Fermi liquid (MFL) behavior in the electron self energy, and linear-in-temperature electrical resistivity. Including particle-hole asymmetry, we find a ``skewed MFL'', with similarities to the skewed non-Fermi liquid of Georges and Mravlje \cite{Georges_Skewed} (see discussion below (\ref{eq:grho})).
We find that the asymmetric thermopower signal in our skewed MFL is as large as that in the non-particle-hole-symmetric complex SYK model \cite{SY92,PGKS,PG99,SS15,Davison16,GKST,Kim_thermo1,Rossini19,Kim_thermo2,Kim_thermo3}. In both cases of Figs.~\ref{fig:kondo} and \ref{fig:ancilla}, and of Section~\ref{kl_vs_single}, the nearly-critical boson action has a non-relativistic form with a linear in time derivative term (see (\ref{eq:boson Greens function})), and it is this feature which results in singular asymmetric contributions to the thermopower near the quantum-critical point. In contrast, there is no such linear in time derivative for symmetry-breaking transitions \cite{Pepin10,QPTbook}, and this is ultimately responsible for their nearly symmetric behavior, and weak thermopower.

We note that our model of purely random spatial interactions might initially seem at odds with the well-established phenomenology of the heavy-fermion compounds, which are famously clean systems, as measured by the residual resistivity. However, recent non-perturbative numerical works by some of us \cite{Patel2023localization,PatelLuntsAlbergo2024} point to the fact that quenched disorder modeled as inhomogeneity in the Yukawa coupling of symmetry-breaking transitions does not necessarily lead to a large residual resistivity. This is the assumption we make here as well, which is why in Sec. \ref{sec:Seebeck} we take the single-particle elastic scattering rate to be small, even though the disorder in the Yukawa coupling is large.

\subsection{Kondo lattice and single-band models}
\label{kl_vs_single}

It is a relatively simple matter to obtain a FL-to-FL* transition in a Kondo lattice model \cite{SVS03,SVS04} as illustrated in Fig.~\ref{fig:kondo}.
The Kondo lattice will serve as our model for Fermi volume-changing transitions in the heavy-fermion compounds \cite{Steglich10,Nakatsuji12,Park24,Coleman01,Shishido05,Steglich09,park2006hidden,Analytis22,Park23}.

It is a far more subtle matter to obtain Fermi-volume-changing transitions without symmetry breaking in single-band electron models. Such transitions are of relevance \cite{Sachdev_Zaanen} to studies of the crossover from the pseudogap to the Fermi liquid in the intermediate temperature regime of the cuprates \cite{Collignon21,Taillefer_Seebeck_PRX_2022,Hoffman14,Seamus14,Taillefer16,Collignon17,Ramshaw22,Yamaji24}, and also in organic compounds \cite{Kanoda23}. We note that our main computations for thermopower in Sections~\ref{sec:theory} and \ref{sec:Seebeck} are carried out using a two-band Kondo lattice model. These results can be directly applied to heavy-fermion compounds. Readers not interested in the extension to the single-band model appropriate for the cuprates may skip over the associated discussion.

We employ the ancilla framework \cite{ZhangSachdev_ancilla} (see also Ref.~\cite{Zou20}) to describe single-band models. This framework uses a bilayer of ancilla qubits, and its main ideas are illustrated in Fig.~\ref{fig:ancilla}.
It is argued that the ancilla qubits can be decoupled by a canonical transformation, leaving behind a single-band Hubbard-like model for the $c_\alpha$ electrons  (see (\ref{Uvalue}) and Appendix~\ref{app:ancilla}). 
As illustrated in Fig.~\ref{fig:ancilla}, we obtain a FL* phase when the bottom ancilla layer forms a spin liquid, while the top two layers combine as in a Kondo lattice to realize a small Fermi volume of Fermi liquid-like quasiparticles; such a FL* description of the pseudogap agrees well with photoemission experiments \cite{Mascot22}. On the other hand, the conventional FL phase is obtained when the two ancilla layers form a trivial rung-singlet state. Note the inversion between Figs.~\ref{fig:kondo} and \ref{fig:ancilla} which plays a key role in our analysis: the FL* phase appears for small Kondo coupling $J_K$ in the Kondo lattice in Fig.~\ref{fig:kondo}, while the FL* phase appears for large $J_K$ in the single-band model in Fig.~\ref{fig:ancilla}.

We note in passing that there is a second framework for describing a FL*-FL transition in single-band models \cite{SSST19}, which is linked by duality in the insulating limit \cite{DQCP3} to the ancilla framework.
This second framework employs bosonic spinons and spinless fermionic holons in the pseudogap phase \cite{sdw09,DCSS15b,DCSS15,CSS17,WuScheurer1,Scheurer:2017jcp,Sachdev:2018ddg,SSST19,WuScheurer2,Bonetti22,Bonetti23}, and there are encouraging comparisons within the pseudogap to experimental data \cite{HeScheurer19} and numerical studies \cite{WuScheurer1,Scheurer:2017jcp,WuScheurer2}.
However, the mechanism for thermopower enhancement across the Fermi-volume-changing transition described in the present paper does not apply to this second framework: the critical Higgs boson does not carry an electromagnetic charge, and consequently the crucial $-i \omega$ term in the boson propagator in (\ref{eq:boson Greens function}) is absent. Therefore, our results lend support to the FL* ancilla description of the pseudogap transition in the cuprates, as does the recent observation of the Yamaji effect \cite{Yamaji24,Sachdev_Zaanen}.

As we show in Fig.~\ref{fig:ancilla_2}, the transition in Kondo lattice models is associated with the condensation of a Higgs boson $b$ in the FL phase.
In contrast, the ancilla transition in single-band Hubbard model has a similar Higgs boson $\Phi$ condensed on the FL* side (see Appendix~\ref{app:ancilla}). 
These features are responsible for one of our main results: the thermopower asymmetry is {\it inverted} between the Kondo lattice and single-band models. As we discuss below, the computed asymmetry of the thermopower is in good accord with that observed in the Kondo lattice system CeRhIn$_5$ \cite{Park24}, which is inverted to that observed in the cuprate La$_{1.6-x}$Nd$_{0.4}$Sr$_x$CuO$_4$ \cite{Collignon21,Taillefer_Seebeck_PRX_2022} (presumed to be described by a single band model).

Thermopower computations in the model of Section~\ref{sec:theory} appear in Section~\ref{sec:Seebeck}, along with a comparison to the experimental observations. We conclude in Section~\ref{sec:conc} with a summary and discussion of implications.

\section{Theory}
\label{sec:theory}

Our computations are carried out for the Kondo Lattice Hamiltonian from the theory of Ref.~\cite{Aldape2022}, given by
\begin{align}
\nonumber
H &= \sum_{\rho = \{c_{\sigma},f_{\sigma},b\}} H_{\rho} + H_{\text{int}}
\\ \nonumber
H_{\rho} &= \sum_{i = 1}^N \sum_k (\epsilon_{\rho,k} - \mu_{\rho}) \rho^{\dag}_{k,i}
\rho_{k,i},
\\ \nonumber
H_{\text{int}} &= \frac{1}{N} \sum_{i,j,l = 1}^N \sum_{r,\sigma} (g'_{ijl}(r) \, c^{\dag}_{r,i,\sigma} f_{r,j,\sigma} b_{r,l} + \text{H.c.}),
\\
& \sum_{i=1}^N (b^{\dag}_{r,i}b_{r,i} - \sum_{\sigma} f^{\dag}_{r,i,\sigma}f_{r,i,\sigma}) = N \kappa\,,
\label{eq:Hamiltonian}
\end{align}
where $c_{\sigma}, f_\sigma$ are conduction and $f$ electrons, and $\sigma$ is a spin index. An auxiliary index $i$ has been introduced for facilitating a large $N$ expansion. The $b$ are (Higgs) bosons mediating the hybridization between the $c$ and $f$ electrons.
The $g'_{ijl}(r)$ are complex Gaussian random variables with variance
\begin{equation}
    \overline{g'_{ijl}(r) g'_{i'j'l'}(r')} = g'^2 \delta_{r,r'} \delta_{i,i'} \delta_{j,j'} \delta_{l,l'},
\end{equation}
which leads to a spatially disordered interaction. This interaction is the result of a Hubbard-Stratonovich decoupling of the original (spatially disordered) Kondo interaction $J_K$. The dispersion of itinerant electrons $\epsilon_{c,k}$ is a generalized tight-binding one for the square lattice and that of the bosons is taken to be quadratic:
\begin{align}
    \nonumber
    \epsilon_{c,k} &= -2 \, t \, (\cos(k_x)+\cos(k_y)) - 4 \, t' \cos(k_x)\cos(k_y) \\
    &~~~~~~~~- 2 \, t'' (\cos(2k_x)+\cos(2k_y)), \nonumber
    \\ 
    \epsilon_{b,k} &= \frac{k^2}{2m_b}.
    \label{eq:dispersions c and b general}
\end{align}

To be concrete, for the hopping amplitudes $t,t',t''$ and chemical potential $\mu_c$ we use the experimental values measured for the cuprate of interest La$_{1.6-x}$Nd$_{0.4}$Sr$_x$CuO$_4$, which are given in Ref. \cite{Collignon21,Taillefer_Seebeck_PRX_2022} (see Appendix~\ref{sec:dispersion}). The precise details of the dispersion for the itinerant $c$ electrons is not important for the results we obtain, and we therefore use the same dispersion when modeling the transition both for the Kondo lattice and single-band cases. The dispersion of the $f$-electrons, on the other hand, is expected to be very different between the two physical systems. In the case of the Kondo lattice model we take them to be heavy, with a simple quadratic dispersion (c.f. Sec. \ref{sec:thermopower Kondo lattice}). In the single-band model we take it to be the same as the dispersion of the $c$ electrons (it can be computed by a mean-field calculation of the ancilla theory \cite{ZhangSachdev_ancilla} in the pseudogap phase, although the exact form is not crucial for the results of this paper; we note that the thermopower is not affected by any kind of ``nesting" between the two Fermi surfaces, as their contributions are separate, and therefore this simplification does not lead to any anomalous features of the Seebeck coefficient). Our computations are carried out in two spatial dimensions. The two-dimensionality is important for the bosonic sector, but the results remain similar if the fermionic sector is three-dimensional (as is the case of YbRh$_2$Si$_2$ \cite{Steglich00}).

The constraint parameter $\kappa$ is what tunes the theory across the QCP, whose critical value is denoted as $\kappa_c$ and the deviation defined as $\Delta \kappa \equiv \kappa - \kappa_c$. For $\Delta \kappa > 0$, the boson $b$ is condensed, while for $\Delta \kappa < 0$ it remains uncondensed. The resulting physical scenarios for the Kondo lattice and single-band models are illustrated in Fig. \ref{fig:ancilla_2}.

\subsection{Propagators at large-$N$}
\label{sec: propagators}

We work in the large $N$ limit, in which the theory is controlled by a self-consistent one-loop saddle point. At this saddle point, the boson Green's function is given by
\begin{equation}
    G_b(i\omega, \mathbf{k}) = \frac{1}{-i \omega + \mathbf{k}^2/(2m_b) + \gamma \abs{\omega} + \Delta_b(T)}.
    \label{eq:boson Greens function}
\end{equation}
Note the $- i \omega$ term, which will ultimately be responsible for the singular asymmetry in the fermion self energies near the quantum critical point, and hence the large thermopower. Such a term is natural for the hybridization (`slave') canonical boson in the Kondo lattice, but is absent for bosonic fields (which are not canonical) representing symmetry-breaking order parameters which are not electrically charged. For complex order parameters involving incommesurate spin/charge density wave orders, the $- i \omega$ term is absent by inversion symmetry.  
The $-i \omega$ term was not present in the analyses of Refs.~\cite{Patel2023universal,Patel2023localization,Li:2024kxr,Hardy24,PatelLuntsAlbergo2024}; for the quantities computed in these earlier papers, the $-i \omega$ term does not lead to any qualitative changes, but will modify various numerical co-efficients. But, as we will see below, it does lead to strong effects in the thermopower.

The co-efficient of the $-i \omega$ term in Eq.~(\ref{eq:boson Greens function}) is unity, because we are working with canonical bosons in our formulation of the theory of the Kondo model. In theories of the Anderson model, the corresponding bosonic field does have a unit co-efficient for the $-i \omega$ term; but we can rescale $b$ to make it unity, leading to changes in the other parameters associated with $b$ in the underlying model.

For the boson, the self-energy correction from the fermions leads to a Landau damping term, with a coefficient given by
\begin{align}
    \gamma = \frac{g'^2 \nu_c \nu_f}{2\pi}\,,
\end{align}
where $\nu_{c/f}$ are the density of states at the Fermi level of the $c$ and $f$ fermions, respectively. The boson gap $\Delta_b(T)$ is obtained by solving the self-consistent equation for $z \equiv {\Delta_b(T)}/{T}$:
\begin{equation}
\begin{split}
    & 2 \pi^2 \frac{\Delta \kappa}{T m_b} \\ & = 
    \int_0^{\infty} \frac{dx}{e^x-1} \[\arctan\(\frac{\gamma \, x}{z-x}\) + \arctan\(\frac{\gamma \, x}{z+x}\)\]
    \\ & 
    - \pi \log(1-e^{-z}) - \frac{\gamma}{1+\gamma^2} z \log\(\frac{\Lambda e}{z T}\),
\end{split}
\label{eq:self-consistent boson gap equation}
\end{equation}
where $\L$ is a UV cutoff scale that comes from the boson bandwidth that we take to be large: $\Lambda/T \gg 1$.

The retarded fermion self-energies written in the variables $x = \omega/T, z = \Delta_b/T$ are given by (see Appendix~\ref{sec:SE})
\begin{equation}
\begin{split}
    & \Sigma_{\rho,R}(T,x,z) = \gamma \frac{T m_b}{2 \nu_{\rho}} \Bigg[-\frac{x}{\pi} \log \left(\frac{\Lambda }{2 \pi  T}\right) - (-1)^{s_{\rho}} i \frac{x}{2} 
    \\ & ~~~~~~
    - i \log \left(\frac{\pi \,  
   \text{csch}\left(\frac{z}{2}\right)}{\Gamma \left(\f12 - i\frac{x-(-1)^{s_{\rho}}z}{2 \pi} \right)^2}\right)\Bigg] + C_{\rho}(T),
\end{split}
   \label{eq:Sigma_c,f full}
\end{equation}
where $s_{c} = 0, s_{f} = 1$, and $C_{\rho}(T)$ is a real frequency-independent term, which can be absorbed into the chemical potential. Importantly, Eq. (\ref{eq:Sigma_c,f full}) is given for the case of $\gamma \ll 1$. The reason for taking this limiting behavior is that it leads to the ``skewed" nature \cite{Georges_Skewed,Aldape2022} of the scattering rates (see below). From here on, we work in this parameter regime. 

Both fermions exhibit a marginal Fermi liquid (MFL) scaling of their self-energies with $T,\omega$. The total scattering rate $\Gamma_{\rho}(T,\omega)$ is given by
\begin{equation}
    \Gamma_{\rho,\text{tot}} = \Gamma_{\rho} - 2 \, \text{Im} \Sigma_{\rho}(x,z,T) = 
    \Gamma_{\rho} + \lambda_{\rho} \, \pi \, T \, g_{\rho}(x,z)\,.
\end{equation}
Here $\Gamma_{\rho}$ is the $T$-independent elastic scattering, which arises from a static random potential (not shown explicitly in Eq.~(\ref{eq:Hamiltonian})); the influence of the random interaction $g'$ is included in the $\Sigma_\rho$, and this yields an inelastic electronic scattering which vanishes as $T \rightarrow 0$ \cite{Patel2023universal}. 
We also have
$x = \o/T, z = \Delta_b/T$, $\lambda_{\rho} = \frac{\gamma m_b}{\pi \nu_{\rho}}$ is an interaction scale, and
\begin{equation}
    g_{\rho}(x,z) = (-1)^{s_{\rho}}\frac{x}{2} + \text{Re}
    \left[\log \left(\frac{\pi \,  
   \text{csch}\left(\frac{z}{2}\right)}{\Gamma \left(\f12 + i\frac{(-1)^{s_{\rho}}z-x}{2 \pi} \right)^2}\right)\right]
   \label{eq:grho}
\end{equation}
is a dimensionless function. Crucially, its form dictates the presence of singular particle-hole asymmetry in both MFLs: $g_{\rho}(x,z)$ contains terms that are both odd and even in $x$. We note in passing that at higher order in $1/N$ the form of $g_{\rho}(x,z)$ will be modified, but the presence of these crucial odd-in-$x$ terms will remain. Following  Georges and Mravlje \cite{Georges_Skewed}, we call this electron state a ``skewed" MFL. Comparing to the analogous function $g(x)$ of Ref.~\cite{Georges_Skewed}, $g_{\rho}(x,z)$ has a more complicated structure due to the $T$ dependence of the mass gap ratio $z(T)$ (which is only a logarithmic $T$ dependence in the quantum-critical region). Ref.~\cite{Georges_Skewed} analyzed a theory exactly at a critical point with conformal symmetry, building on the large particle-hole asymmetry in the Sachdev-Ye-Kitaev model \cite{SY92,PGKS,PG99,SS15,Davison16,GKST,Kim_thermo1,Rossini19,Kim_thermo2,Kim_thermo3}, that has a $T$-independent $z(T)$. 

To a good approximation, the renormalized quasiparticle weights $Z_{\rho}(T,\omega)$ can be estimated as energy-independent, $Z_{\rho}(T,\omega) \approx Z_{\rho}(T,\omega = 0)$ \cite{Georges_Skewed}. They are then computed as
\begin{equation}
\begin{split}
    \frac{1}{Z_{\rho}(T)} &= 1 - \frac{1}{T}
    \frac{\partial \text{Re} \Sigma_{\rho}(x,T)}{\partial x} \Bigr|_{\substack{x=0}}
    \\ & 
    = 1 + \frac{\lambda_{\rho}}{2} 
    \[ \log \left(\frac{\Lambda }{2 \pi  T}\right) - \text{Re} \, \psi\(\frac{1}{2} + i \frac{z}{2\pi}\)
    \],
\end{split}
\label{eq:Z inverse}
\end{equation}
where $\psi(x)$ is the digamma function. This logarithmic form is what is expected for a MFL. The second term involving the digamma function is essentially constant in temperature, as can be confirmed numerically. 

\section{Seebeck coefficient}
\label{sec:Seebeck}

The thermopower is given by ($e$ here denotes electric charge)
\begin{align}
    S = - \frac{L_1}{e \, L_0}\,,
\end{align} 
with the $L_n$ are called Onsager coefficients and are given by
\begin{align}
    L_0 &= -\lim_{\omega\to 0} \frac{\text{Im} \, \Pi_{J J}(\omega)}{\omega}
    \\
    L_1 &= - \frac{1}{T} \lim_{\omega\to 0} \frac{\text{Im} \, \Pi_{J_Q J}(\omega)}{\omega},
\end{align}
where $J$ is the electrical current and $J_Q$ is the heat current. Each current is the sum of currents from all three species: $J = \sum_{\rho = c, f, b} J^{(\rho)}, J_Q = \sum_{\rho = c, f, b} J^{(\rho)}_Q$. From here on we set $k_B = e = 1$ (except when we restore them in final results).

Importantly, in this model, to the order in $1/N$ we are working, all cross-current correlators vanish. Therefore, the correlators reduce to the sum over single species ones, and so do the $L_n$,
\begin{align}
    L_0^{(\rho)} &= - \s_{\rho} \lim_{\omega\to 0} \frac{\text{Im} \, \Pi_{J^{(\rho)} J^{(\rho)}}(\omega)}{\omega}
    \\
    L_1^{(\rho)} &= - \frac{1}{T} \s_{\rho} \lim_{\omega\to 0} \frac{\text{Im} \, \Pi_{J_Q^{(\rho)} J^{(\rho)}}(\omega)}{\omega},
\end{align}
where $\s_{\rho}$ is the spin degeneracy of the species: $\s_{c,f} = 2, \s_{b} = 1$. Using the notation of Ref. \cite{Georges_Skewed}, we can write them as
\begin{equation}
    L_n^{(\rho)} = \frac{1}{T^n} \int d \omega \(- \frac{\partial f_{\rho}}{\partial \omega}\) \omega^n \; \mathcal{T}^{(\rho)}(\omega),
    \label{eq:L_n general}
\end{equation}
where $f_{\rho}$ is the occupation function of each species, and
\begin{equation}
\begin{split}
    \mathcal{T}^{(\rho)}(\omega) & = \s_{\rho} \pi \int \frac{d^2 k}{(2 \pi)^2} v_{\rho,\mathbf{k}}^2 \, A_{\rho}(\mathbf k, \omega)^2 
    \\ & = \pi \int d \epsilon \, \Phi(\epsilon) \, A_{\rho}(\epsilon, \omega)^2,
\end{split}    
    \label{eq:T general}
\end{equation}
where
\begin{equation}
\Phi(\epsilon) = \s_{\rho} \int \frac{d^2 k}{(2\pi)^2} v_{\rho,\mathbf{k}}^2 \delta(\epsilon - \epsilon_{\mathbf k}^{(\rho)}).
\end{equation}
is the transport function and $A_{\rho}(\mathbf k, \omega)$ is the spectral function. 

The emergent $U(1)$ gauge field that couples $f$ and $b$ will act to renormalize the current vertices. In the calculation of the electrical conductivity this leads to exact Ioffe-Larkin constraints on the current correlators, giving a combination of in-series and in-parallel current additions. 

Although both electrical and thermal current vertices should get renormalized, at the order of $1/N$ we are working at, the renormalizations to the thermal current vertex cancel; see Appendix~\ref{sec:vertices}. This leads to a convoluted relationship of the currents of the different species, which cannot be simply summarized as in-parallel or in-series addition. The Onsager coefficients are written as 
\begin{align}
    L_0 &= L_0^{(c)} 
    + \(\(L_0^{(f)}\)^{-1} + \(L_0^{(b)}\)^{-1}\)^{-1}
    \\
    L_1 &= 
    L_1^{(c)} - \frac{L_1^{(f)} L_0^{(b)} + L_1^{(b)} L_0^{(f)}}{L_0^{(f)} + L_0^{(b)}},
    \label{eq:L_0 and L_1}
\end{align}
and the total thermopower becomes
\begin{equation}
    S = - \frac{L_1}{L_0} 
    = - 
    \frac{\displaystyle L_1^{(c)} - \frac{L_1^{(f)} L_0^{(b)} + L_1^{(b)} L_0^{(f)}}{L_0^{(f)} + L_0^{(b)}}}    
    {\displaystyle L_0^{(c)} 
    + \frac{L_0^{(f)} L_0^{(b)}}{L_0^{(f)} + L_0^{(b)}}}.
    \label{eq:Seebeck general expression}
\end{equation}
Again, these equations only come from the renormalization of the electric charge.

The Onsager coefficients of fermions and bosons are analyzed differently, due to the presence/absence of a Fermi surface. For the two fermion species, $\rho = c, f$, we can use the results of Ref. \cite{Georges_Skewed}, as their assumptions apply to our model. These assumptions are that (i) the scattering rate (both elastic and inelastic) is momentum independent, and (ii) the self-energy is small compared to the bare terms. We can write
\begin{align}
    L_0^{(\rho)} & = \Phi_0^{(\rho)} \langle \, 
    \tau_+(T, x T)
    \rangle + \frac{T}{Z(T)} \Phi_0^{(\rho)'} \langle x \, \tau_-(T, x T) \rangle,
    \\
    L_1^{(\rho)} & = \Phi_0^{(\rho)} \langle \, 
    x \, \tau_-(T, x T)
    \rangle + \frac{T}{Z(T)} \Phi_0^{(\rho)'} \langle x^2 \, \tau_+(T, x T) \rangle. 
    \label{eq:fermion L_n}
\end{align}
where we have additionally assumed that the quasiparticle residue $Z(T,\omega)$ is energy-independent (in practice it indeed depends only weakly on the energy near the Fermi surface). Here, $\Phi_0 = \Phi(\eps_F), \Phi'_0 = \Phi'(\eps_F)$. The frequency-dependent scattering time $\tau(T,\omega) = \frac{1}{\Gamma_{\text{tot}}(T,\omega)}$ is divided into even and odd components: $\tau_{\pm}(T,\omega) \equiv \f12 (\tau(T,\omega) \pm \tau(T,-\omega))$. The brackets notation is
\begin{equation}
    \langle F(x) \rangle \equiv \int_{-\infty}^{\infty} dx \frac{F(x)}{4 \cosh^2(\frac{x}{2})}.
\end{equation}

In the case of the boson Onsager coefficients, due to the absence of a large scale $\eps_F \gg T$, we cannot use Eqs. (\ref{eq:fermion L_n}). However, due to the completely disordered nature of the interactions the scattering rate is $\mathbf{k}$-independent. We can therefore still use the expression for $\mathcal{T}(\omega)$ from Eq. (\ref{eq:T general}).

\subsection{Onsager coefficients}

We are now in a position to compute the Onsager coefficients at the leading order in $1/N$ for the theory in Eq. (\ref{eq:Hamiltonian}). The elastic scattering components, $\Gamma_{\rho}$, are very important, as they set the temperature scales for high- and low-temperature asymptotic behavior. These crossover temperature scales, $T^*_{\rho}$, are defined by comparing the elastic and inelastic scattering components, and are implicitly given by the solution to 
\begin{equation}
    T^*_{\rho} = \frac{\Gamma_{\rho}}{\lambda_{\rho} \, \pi \,  g(0,z(T^*_{\rho}))}.
    \label{eq:T star both fermions}
\end{equation}
Using temperature variables that are scaled with respect to the crossover temperatures, $\theta_{\rho} = T/T^*_{\rho}$, the individual Onsager coefficients for the fermions are summarily written as
\begin{align}
    & L_n^{(\rho)} = \frac{\Phi^{(\rho)}_0}{\Gamma_\rho} \Bigg[F^{(\rho)}_n(\theta_\rho) 
    \\ & 
    + \eta_\rho \, \theta_\rho \( \frac{T^*_\rho}{\Gamma_\rho} + \frac{\log \left(\frac{\Lambda }{2 \pi  T}\right) - \text{Re} \, \psi\(\frac{1}{2} + i \frac{z}{2\pi}\)}{2\pi g_{\rho}(0,z(T^*_{\rho}))} \) F^{(\rho)}_{n+1}(\theta_\rho) \Bigg], \nonumber 
\end{align}
where $\eta_{\rho} = \Gamma_{\rho} \Phi'^{(\rho)}_0/\Phi^{(\rho)}_0$ is the ratio of the elastic scattering rates to the characteristic energy scales associated with the band-structure asymmetries, and 
\begin{equation}
    F_n^{(\rho)}(\theta_{\rho}) \equiv \left<\frac{x^n}{1+ \displaystyle \theta_{\rho} \frac{g_{\rho}(x,z)}{g_{\rho}(0,z(T^*_{\rho}))}}\right>.
    \label{eq:F functions for c fermions}
\end{equation}

For the bosonic coefficients $L_n^{(b)}$ we need the boson transport function,
\begin{equation}
    \Phi^{(b)}(\epsilon) = \int \frac{d^2 k}{(2 \pi)^2} v_{\mathbf{k}}^2 \, \delta(\epsilon - \epsilon^{(b)}_{\mathbf k})
    = \frac{1}{\pi} (\epsilon - \Delta_b) \, \Theta(\epsilon - \Delta_b).
\end{equation}
Inserting the spectral function computed from Eq. (\ref{eq:boson Greens function}) into Eq. (\ref{eq:L_n general}), the final expression is given by
\begin{equation}
\begin{split}
    L_n^{(b)} = \frac{1}{2 \pi^2} \int_{-\infty}^{\infty} & d x \, \frac{x^n}{4 \sinh^2(\frac{x}{2})} \,
    \\ & 
    \(1 + \frac{(x - z) \(\frac{\pi}{2} + \arctan{\frac{x - z}{\gamma \abs{x}}}\)}{\gamma \abs{x}}\),
\end{split}    
    \label{eq:L_n boson}
\end{equation}
where, as before, $x = \omega/T$ and $z = \Delta_b/T$. Since the temperature dependence of $z(T)$ is very weak for all but the smallest temperatures, the coefficients $L_n^{(b)}$ are largely temperature-independent in the temperature regimes that we investigate. Importantly, the boson spectral function does not contain an elastic scattering term, as the boson does not interact with the static impurities in the system.



\subsection{Seebeck coefficient for the Kondo lattice}
\label{sec:thermopower Kondo lattice}

\begin{figure*}
    \centering
    \includegraphics[width=0.4\linewidth]{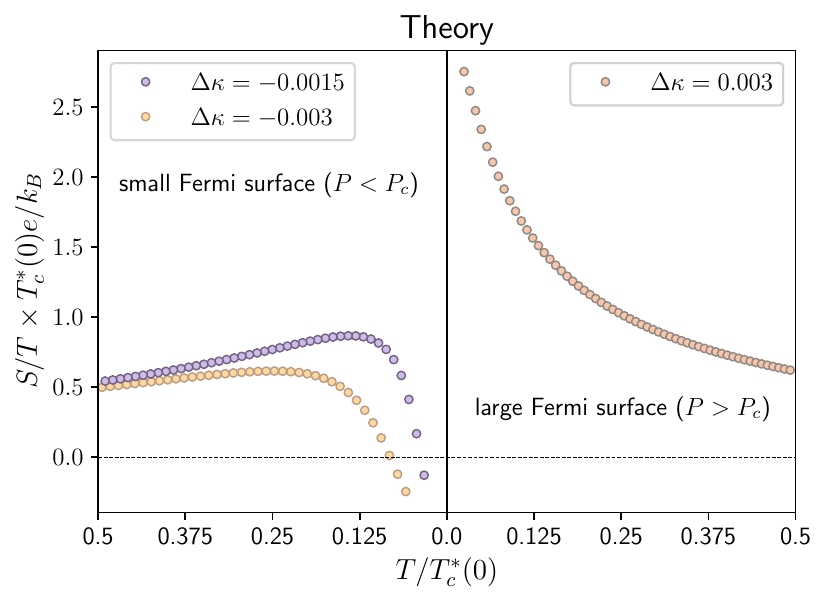}
    ~~~~
    \includegraphics[width=0.3\linewidth]{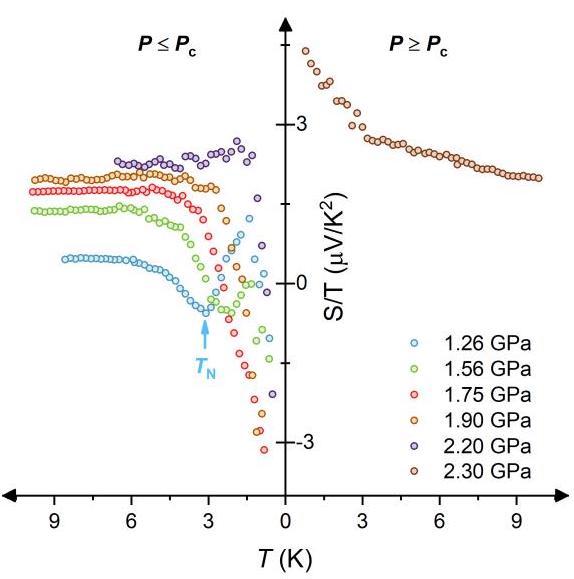}    
    \caption{(a) Seebeck coefficient for small temperatures on the two sides of the transition, $\Delta \kappa \lessgtr 0$. For the Kondo lattice case relevant for the heavy fermions, we choose the microscopic parameter values as (c.f. Sec. \ref{sec:thermopower Kondo lattice}) $\Gamma_c = \Gamma_f = 0.02$, $\Lambda = 100$, $\gamma = 0.05$, $m_b = 25$, $\nu_c = 0.3054$, $\Phi^{(c)}_0 =  1.064$, $\Phi'^{(c)}_0 = 0.83$, $\nu_f = 3$, $\Phi^{(f)}_0 = 0.047$, $\Phi'^{(f)}_0 = 1/\pi$.
    We normalize the temperature axis by the crossover scale at criticality, $T_c^*(\Delta \kappa = 0)$. The Seebeck coefficient itself is multiplied by $T_c^*(\Delta \kappa = 0) e/k_B$, giving a dimensionless quantity. The product of the values on the $x$ and $y$ axes gives the thermopower in units of $k_B/e$, and is normally less than one in real materials. We can see in our theoretical modeling this holds as well. (b) The experimental value of $S/T$ in $\text{CeRhIn}_5$, taken from Ref. \cite{Park24}. Looking at the three values of pressure that are closest to the critical value (on either side of it), $P = 1.90 ~\text{GPa}, 2.20 ~\text{GPa}, 2.30 ~\text{GPa}$, all the qualitative features of the curves match up with those in (a).}
    \label{fig:S over T HF}
\end{figure*}

With all of the Onsager coefficients in hand, we can use Eq. (\ref{eq:Seebeck general expression}) to model the Seebeck coefficient, $S/T$, across the Fermi-volume changing transition in heavy fermion compounds. For our model parameters that have not been specified in Sec. \ref{sec:theory}, we choose the following. We set the Landau damping parameter to $\gamma = 0.05$ and the boson UV cutoff $\Lambda = 100$. As mentioned in Sec. \ref{sec:theory}, we take the dispersion of the $f$-electrons to be quadratic $\epsilon_f(k) = \frac{k^2}{2 m_f} - \mu_f$. The mass is chosen to be $m_f = 10 m_c$, and the chemical potential is chosen to make the density of the two fermions approximately equal (motivated by stoichiometric considerations in the real material). These choices give $\nu_f = 3, \Phi_0^{(f)} = 0.047, \Phi_0^{'(f)} = 1/\pi$. The boson mass is chosen to be $m_b = 25 \approx 1.2 (m_f + m_c)$ (as it should be $m_b \gtrsim m_c + m_f$). We take the elastic scattering rate for the $c$-electron to be small, $\Gamma_c = 0.02$. This is a good assumption, as heavy fermion compounds are usually very clean as measured by the residual resistivity, and the electrical conductivity in this model comes entirely from the $c$-electron \cite{Aldape2022}. The elastic scattering rate for the $f$-electrons is then larger by a factor of $\nu_f/\nu_c \approx 10$, so we take $\Gamma_f = 0.2$.  Inserting all of these necessary components, the resulting Seebeck coefficient is shown in Fig. \ref{fig:S over T HF}.
We can see that the Seebeck coefficient (i) is of the opposite sign as would be determined by the band structure alone, and initially increases in magnitude with decreasing temperature on both sides of the transition, both as expected for a `skewed' MFL, (ii) is larger on the `large Fermi surface' side of the transition and (iii) has qualitatively different behavior with $T$ on the two sides of the transition. 

The reason for this behavior is clarified using a small-temperature expansion of the Seebeck coefficient, analyzed below in Sec. \ref{sec:low-T analytical expansion}. One key takeaway is that the behavior of $S/T$ is dominated by the $c$-electrons on both sides of the transition, due to the small elastic scattering rates. The degree of `skewness' is wrapped up in a single coefficient, which determines the temperature scale at which the sign of $S/T$ is restored to the non-interacting one. This temperature scale turns out to be quite sensitive to the sign of $\Delta \kappa$, so that in the temperature range plotted in Fig. \ref{fig:S over T HF} the small Fermi surface side ($\Delta \kappa < 0$) shows this change in sign, while the large Fermi surface side ($\Delta \kappa > 0$) does not even show a downturn (at a low enough temperature this downturn will occur). This is an additional manifestation of the sensitivity of the Seebeck coefficient to the degree of particle-hole asymmetry in the system. 

In order to facilitate a qualitative comparison with experimental measurement, in Fig. \ref{fig:S over T HF}b we show a plot of $S/T$ in $\text{CeRhIn}_5$ across the pressure-tuned Fermi-volume changing transition, taken directly from Ref. \cite{Park24}. Focusing on the three curves closest to the transition, we can see that our theoretical predictions are qualitatively very similar to the experimental behavior. This is suggestive of the fact that the metallic state inside of the critical fan of the transition is indeed a ``skewed" MFL \cite{Georges_Skewed}.  

\subsection{Seebeck coefficient for the single band model}
\label{sec:thermopower single band cuprate}

\begin{figure*}
    \centering
    \includegraphics[width=0.4\linewidth]{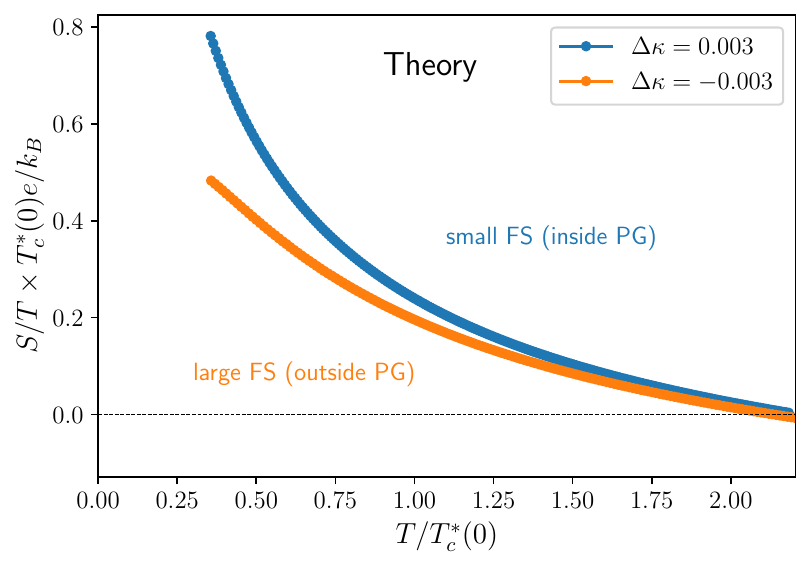}
    ~~~~
    \includegraphics[width=0.4\linewidth]{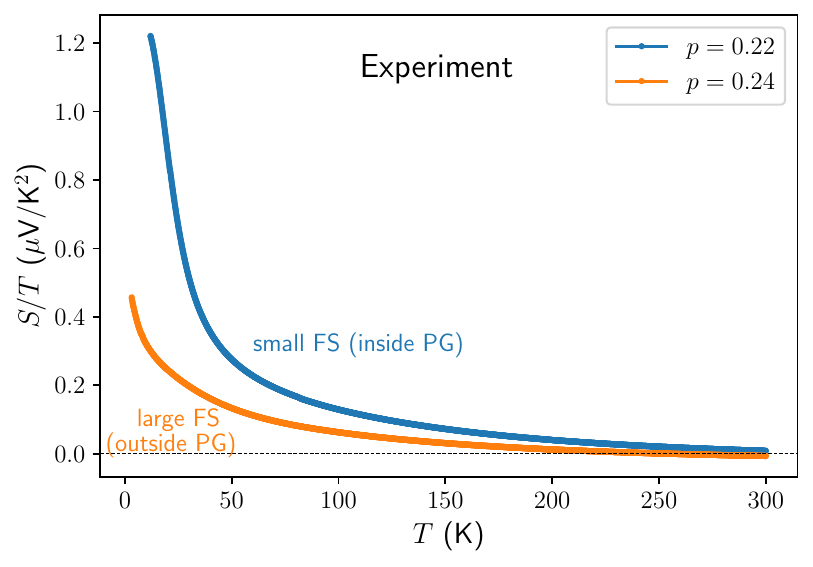}
    \caption{(a) Seebeck coefficient for small temperatures on the two sides of the transition, $\Delta \kappa \lessgtr 0$. For the single-band case relevant for the cuprates, we choose the microscopic parameter values as (c.f. Sec. \ref{sec:thermopower single band cuprate}) $\Gamma_c = \Gamma_f = 0.02$, $\Lambda = 100$, $\gamma = 0.05$, $m_b = 5$, $\nu_c = \nu_f = 0.3054, ~ \Phi^{(c)}_0 = \Phi^{(f)}_0 = 1.064, ~ \Phi'^{(c)}_0 = \Phi'^{(f)}_0 = 0.83$. The axes are normalized as in Fig. \ref{fig:S over T HF}a. As for the parameters modeling the heavy-fermion compounds, the numerical value of thermopower behaves as expected (cf. the caption of Fig. \ref{fig:S over T HF}). (b) The experimental value of the in-plane Seebeck coefficient in La$_{1.6-x}$Nd$_{0.4}$Sr$_x$CuO$_4$ on both sides of the pseudogap critical point doping value, reproduced from Ref. \cite{Taillefer_Seebeck_PRX_2022} with permission from the authors. The qualitative features of the curves match up with those in (a).}
    \label{fig:S over T cuprates}
\end{figure*}

Using the alternative view of Eq. (\ref{eq:Hamiltonian}) as coming from the ancilla theory description of the single band transition, we can also make a comparison to thermopower measured in cuprates \cite{Collignon21,Taillefer_Seebeck_PRX_2022}. For this, we modify the dispersion of the $f$ electrons by making it identical to that of the $c$-electrons, as explained in Sec. \ref{sec:theory}. The boson mass is taken to be $m_b = 5 \approx 1.3 (m_f + m_c)$, and all other parameters are left the same as in Sec. \ref{sec:thermopower Kondo lattice}. The resulting curves are shown in Fig. \ref{fig:S over T cuprates}, next to a reproduction of the data from Ref. \cite{Taillefer_Seebeck_PRX_2022}. 
We can see that the behavior as a function of $T$ and $\Delta \kappa$ again qualitatively reproduces the behavior as a function of $T$ and the doping value relative to the critical one. We take this as support for the pseudogap critical point being governed by a non-symmetry-breaking Fermi-volume-changing transition, in the critical fan of which there exists a ``skewed" MFL state.

\subsection{Small temperature expansion}
\label{sec:low-T analytical expansion}

\begin{figure}
    \centering
    \includegraphics[width=0.97\linewidth]{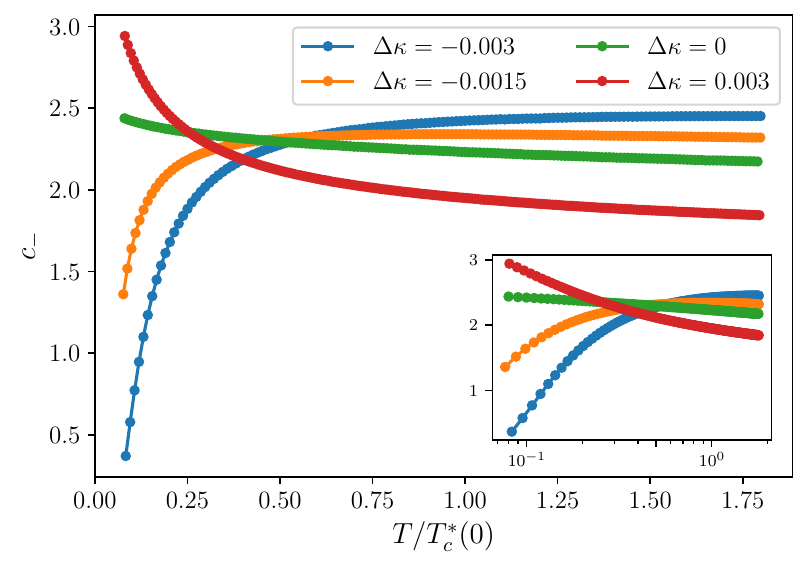}
    \caption{Behavior of $c_{-}(T)$ versus $\theta_c = T/T_c^*(0)$ for various $\Delta \kappa$. The parameter values are taken from Sec. \ref{sec:thermopower single band cuprate}. The inset is the same plot, albeit with a logarithmic scale for the temperature axis, in order to illustrate the asymptotic behavior at the smallest $\theta_c$.}
    \label{fig:c_ vs T}
\end{figure}

To better understand the results of Figs. \ref{fig:S over T HF}, \ref{fig:S over T cuprates}, we analyze the thermopower analytically via a small temperature expansion, specifically $\theta_c \ll 1$ (this implies $\theta_f \ll 1$ for both parameter regimes analyzed above, since $\theta_f \ll \theta_c$ in the Kondo lattice case and $\theta_f \sim \theta_c$ in the single-band case). Here, we only expand the $F^{(\rho)}_n$ functions, i.e. we ignore the $\theta_\rho$-dependence of $z(T)$, which is known to be at most logarithmic in the temperature regime of interest (temperatures above the crossover scale between the quantum and classical critical regimes). Keeping the parameters for the two fermion species distinct for the moment, and restoring $k_B$ and $e$, we write the small $\theta_c$ expression for $S$ as 
\begin{widetext}
\begin{equation}
\begin{split}
    S = - & \frac{k_B}{e}\frac{1}{\Phi^{(c)}_0 
    + \frac{\Phi^{(f)}_0 L_0^{(b)} \Gamma_c}{\Phi^{(f)}_0 + L_0^{(b)} \Gamma_f}}
    \Bigg[
    \theta_c \(\frac{\pi^2}{3} \Gamma_c \, \Phi'^{(c)}_0 \, \( \frac{T^*_c}{\Gamma_c} + \frac{\log \left(\frac{\Lambda }{2 \pi  T}\right) - \text{Re} \, \psi\(\frac{1}{2} + i \frac{z}{2\pi}\)}{2\pi g_c(0,z(T^*_c))} \) - c_-(T) \Phi^{(c)}_0 \) 
    \\ & 
    ~~~- \theta_c \frac{T_c^*}{T_f^*} \frac{L_0^{(b)}\Gamma_c}{\Phi^{(f)}_0 + L_0^{(b)}\Gamma_f} \(\frac{\pi^2}{3} \Gamma_f \, \Phi'^{(f)}_0 \, \( \frac{T^*_f}{\Gamma_f} + \frac{\log \left(\frac{\Lambda }{2 \pi  T}\right) - \text{Re} \, \psi\(\frac{1}{2} + i \frac{z}{2\pi}\)}{2\pi g_f(0,z(T^*_f))} \) + c_-(T) \Phi^{(f)}_0 \)  
    \\ & 
   ~~~ - \frac{\Phi^{(f)}_0 L_1^{(b)} \Gamma_c}{\Phi^{(f)}_0 + L_0^{(b)}\Gamma_f}
    \Bigg]
    + \mathcal{O}(\theta_c^2),
\end{split}
    \label{eq:Seebeck small theta}
\end{equation}
\end{widetext}
where 
\begin{align}
    c_-(T) \equiv \left< x \frac{g_c(x,z)}{g_c(0,z(T^*_c))}\right> = -\left< x \frac{g_f(x,z)}{g_f(0,z(T^*_f))}\right>\,.
\end{align}
To make further progress, we use the fact that the elastic scattering rates are small, $\Gamma_c \ll 1, \Gamma_f \ll 1$ (we do not rely on their approximate equality). This, in conjunction with the fact that $T^*_c \leq T^*_f$ and $L_n^{(b)} \sim \mathcal{O}(1)$ (not shown), implies that we can focus on the first term in Eq. (\ref{eq:Seebeck small theta}) for a large temperature window. Therefore, in the low-intermediate temperature range, the thermopower can be estimated as coming exclusively from the $c$-electrons, which by themselves already form a ``skewed" MFL,
\begin{equation}
\begin{split}
    S \approx - \frac{k_B}{e}
    \theta_c \Bigg[ &
    \frac{\pi^2}{3} \eta_c \,\( \frac{T^*_c}{\Gamma_c} + \frac{\log \left(\frac{\Lambda }{2 \pi  T}\right) - \text{Re} \, \psi\(\frac{1}{2} + i \frac{z}{2\pi}\)}{2\pi g_c(0,z(T^*_c))} \) 
    \\ & 
    - c_-(T)\Bigg].
\end{split}
    \label{eq:Seebeck c electrons small theta}
\end{equation}
The terms in round parentheses multiplying $\eta_c$ are positive and therefore drive $S$ to be of the same sign as that of the free theory, i.e. $\sign(S) = -\sign(\eta_c)$. The coefficient $c_-(T)$ counteracts this effect, and over a large low-temperature range changes the sign of $S$ \cite{Georges_Skewed}. However, $c_-(T)$ itself has some non-trivial $T$-dependence, which depends crucially on the sign of $\Delta \kappa$. To illustrate this, we plot $c_{-}(T)$ for various values of $\Delta \kappa$ using the same model parameters as in Sec. \ref{sec:thermopower single band cuprate} in Fig. \ref{fig:c_ vs T}.
For all values of $\Delta \kappa$, $c_-(T)$ is roughly constant and $\sim \mathcal{O}(1)$ for all but the smallest temperatures. Below $\theta_c \approx 0.5$, $c_-(T)$ vanishes super-logarithmically for $\Delta \kappa < 0$, while growing (slightly) sub-logarithmically for $\Delta \kappa \geq 0$. This explains the behavior of $S/T$ in this low-intermediate temperature range, as its sign is determined by $c_-(T)$. While $c_-(T)$ is roughly constant, $S/T > 0$ and changes very slowly and eventually decays to $S/T < 0$ at larger temperatures. At lower temperatures, for $\Delta \kappa \geq 0$, the sub-logarithmic increase in $c_-(T)$ will counteract the logarithm coming from the MFL quasiparticle weight down to a critical temperature, after which the logarithmic part will take over and restore $\sign(S) \to -\sign(\eta_c)$. This critical temperature will be exponentially suppressed by some power of $\eta_c \ll 1$, which explains its absence from both the theory and experimental plots of Figs. \ref{fig:S over T HF},\ref{fig:S over T cuprates}. However, for $\Delta \kappa < 0$, $c_-(T)$ drops to zero (super-logarithmically), and this sign restoration occurs at a much larger $T$, again explaining the low-temperature downturn of $S/T$ in both theory and experiment in Figs. \ref{fig:S over T HF},\ref{fig:S over T cuprates}. We note that both of these behaviors are not fine-tuned features of the theory. For all values of $\Delta \kappa$ the form of $S/T$ at ultra-low temperatures is still logarithmic, $S/T \sim - \frac{1}{T^*_c}\frac{\pi}{6} \eta_c \log \left(\frac{\Lambda }{2 \pi  T}\right)/g_c(0,z(T^*_c))$, and satisfies $\sign(S) = -\sign(\eta_c)$ (as $g_c(0,z(T^*_c)) > 0$), as in the case studied in Ref. \cite{Georges_Skewed}.  


\section{Discussion}
\label{sec:conc}

In this work, we have computed the thermopower in a Kondo lattice model across the non-symmetry-breaking, Fermi-volume-changing quantum phase transition, using a controlled large-$N$ approach. Our results show that the thermopower is large, and its behavior across the transition is not symmetric, and the side with the larger Fermi surface has an enhanced Seebeck coefficient, whose temperature-dependence is also non-trivial due to the itinerant electrons being a marginal Fermi-liquid. Furthermore, the side with the smaller Fermi surface has a non-monotonic temperature dependence in the low-intermediate temperature range. Comparing our results to those measured in the heavy fermion compound $\text{CeRhIn}_5$ in Ref. \cite{Park24} gives a good qualitative match to the experimental data, as shown in Fig.~\ref{fig:S over T HF}. 

Symmetry-breaking transitions do not have the $-i \omega$ term in the critical boson propagator in
(\ref{eq:boson Greens function}) \cite{QPTbook}, and so their thermopower is quite weak, and mostly symmetric across the transition.
In contrast, the asymmetric thermopower signal in the ``skewed MFL'' we have described here is as large as that in the non-particle-symmetric complex SYK model \cite{SY92,PGKS,PG99,SS15,Davison16,GKST,Kim_thermo1,Rossini19,Kim_thermo2,Kim_thermo3} and the model of Georges and Mravlje \cite{Georges_Skewed}.

Another important point concerns the origin of the asymmetrical behavior in Fig. \ref{fig:S over T HF} and the downturn of $S/T$ at small values of temperature. In this work we investigated the behavior inside the ``quantum critical fan" only. Therefore, the non-monotonicity is not related to the crossover temperature of this quantum critical region, as such a scale is much lower than our smallest temperatures. This is in contrast to Ref. \cite{Pepin10}, where a somewhat similar behavior was argued to exist precisely due to this crossover scale.

Heavy fermion compounds typically have more than the two low-energy bands mentioned in this work, so it may seem strange to not take them into account. In particular, the various elastic scattering rates will determine the sequence of temperature windows where the thermopower is dominated by some bands and not others. However, near the Fermi-volume-changing transition, only one of the itinerant bands hybridizes with the localized $f$ electrons, and only that band will get renormalized into a ``skewed" MFL, while the others remain Fermi-liquid-like. This implies \cite{Georges_Skewed} that at low temperatures the main contribution to the Seebeck coefficient will come from this band only, and our results should still hold.

Interpreting the same model as coming from the ancilla theory of a Fermi-volume changing transition in a single-band model, we also make a comparison to recent experimental results on the thermopower in the cuprate La$_{1.6-x}$Nd$_{0.4}$Sr$_x$CuO$_4$ \cite{Collignon21,Taillefer_Seebeck_PRX_2022}. Our results again qualitatively match the experimental data in a large temperature window (as shown in Fig.~\ref{fig:S over T cuprates}), and reproduce the effect of a larger thermopower on the side of the smaller Fermi surface (pseudogap side of the critical point $p^*$). 
We regard this match as significant support for the existence of an underlying non-symmetry-breaking Fermi-volume-changing transition in the cuprates for the onset of the pseudogap at intermediate temperatures. Furthermore, it is the ancilla inverted Kondo lattice description \cite{ZhangSachdev_ancilla} which is consistent with the observations, as we discussed in Section~\ref{kl_vs_single}.

Another observation that adds support to this idea is the behavior of the Hall coefficient $R_H$ across $p^*$. Experiments \cite{Collignon17} show that $R_H$ drops when going from $p > p^*$ to $p < p^*$. This is consistent with $p^*$ being a Fermi-volume-changing transition, as the carrier density (responsible for Hall transport) is proportional to the Fermi volume. In fact, the behavior of $R_H$ as a function of $p$ and $T$ bears a remarkable resemblance to the corresponding behavior of $S/T$ \cite{Collignon21,Taillefer_Seebeck_PRX_2022}, suggesting a similar mechanism for the two.

In comparing to the observations in the cuprates in Fig. \ref{fig:S over T cuprates}a, we have cut off the smallest temperature dependence, in order to illustate the match between Figs. \ref{fig:S over T cuprates}a and \ref{fig:S over T cuprates}b in the rest of the temperature window. Lower-temperature behavior would eventually show a similar non-monotonicity as in Fig. \ref{fig:S over T HF}a, which is absent in the experimental measurements. Our interpretation of this is that, at a low enough temperature scale, the blue curve in \ref{fig:S over T cuprates}b (large Fermi surface) will eventually reach a maximum and exhibit a (super-logarithmic) downturn. This can be viewed as a prediction of our theory. 

One unrealistic simplification our theory makes is that the interaction between the electrons and the Higgs field is purely random in space, $g'(r)$. A much more reasonable assumption is that the random part is subdominant to a translationaly invariant piece, $g + g'(r)$ with $g \gg g'$. The inclusion of $g$ will complicate the calculations in this paper, but we suspect it will not affect the low-temperature thermoelectric transport properties, as $g'(r)$ was argued to be most relevant for low-temperature electrical transport. However, a confirmation of this assumption from an explicit calculation with both couplings is highly desired.

\section*{Acknowledgments}
We thank Antoine Georges for insightful comments and explaining the results of Ref.~\cite{Georges_Skewed}.
We also thank Ehud Altman, Gael Grissonnanche, and Louis Taillefer for useful discussions. We thank the authors of Ref.~\cite{Taillefer_Seebeck_PRX_2022} for providing us with the data for Fig.~\ref{fig:S over T cuprates}b, and the authors of Ref.~\cite{Park24} for permission to reproduce Fig.~\ref{fig:S over T HF}b. This research was supported by the U.S. National Science Foundation grant No. DMR-2245246, by the Harvard Quantum Initiative Postdoctoral Fellowship in Science and Engineering, and by the Simons Collaboration on Ultra-Quantum Matter which is a grant from the Simons Foundation (651440, S. S.). The Flatiron Institute is a division of the Simons Foundation.

\onecolumngrid
\appendix

\section{Ancilla theory of single-band model}
\label{app:ancilla}

The ancilla theory of Ref.~\cite{ZhangSachdev_ancilla} was outlined in Fig.~\ref{fig:ancilla}. The degrees of freedom are the physical electrons $c_\alpha$ in the top layer, and two layers of ancilla spins, ${\bm S}_1$ and ${\bm S}_2$ in the bottom two layers. Eliminating these spin layers by a canonical ({\it i.e.} Schrieffer-Wolff type)
transformation leads to a Hubbard-type single band Hamiltonian for the electrons in the top layer. This canonical transformation can be made precise at large $J_\perp$ when we obtain only the Hubbard interaction \cite{Maria21}
\begin{align}
    U = \frac{3 J_K^2}{8 J_\perp} + \frac{3 J_K^3}{16 J_\perp^2} + \mathcal{O} (1/J_\perp^3)\,. \label{Uvalue}
\end{align}

We begin by recalling here the nature of the effective Hamiltonian within the FL* phase. The bottom layer of the ${\bm S}_2$ spins is assumed to form a spin liquid with fractionalization and no broken symmetry. An important advantage of the ancilla formulation is that this could be any spin liquid, and not just that associated with a spinon Fermi surface (which was assumed for the $f_\sigma$ spinons in (\ref{eq:Hamiltonian})). In mean-field theory, this ${\bm S}_2$ layer  is decoupled from the top two-layers. For the intermediate ${\bm S}_1$ layer, we express the spin degrees of freedom in terms of a fractionalized fermion $F_a$, as indicated in Fig.~\ref{fig:ancilla}. The index $a$ is {\it not\/} a physical spin index, but that obtained by transforming to a rotating reference frame in spin space \cite{sdw09,ZhangSachdev_ancilla,Bonetti22}. Consequently, $F_a$ is subject to SU(2)$\times$U(1) gauge transformations \cite{ZhangSachdev_ancilla,Zou20,SS_QPM}:
\begin{align}
    F_a \rightarrow U_{ab} F_b \quad, \quad F_a \rightarrow e^{i \theta} F_a\,. \label{su2g1}
\end{align}
As in (\ref{eq:Hamiltonian}), we also have a hybridization Higgs boson $\Phi_{\sigma a}$ replacing the $b$ boson. This boson couples the top two-layers by a Yukawa coupling, similar to $H_{\rm int}$ in (\ref{eq:Hamiltonian}):
\begin{align}
    H_{\rm int} \sim c_{\sigma}^\dagger \Phi_{\sigma a} F_a \label{Hint}
\end{align}
The gauge transformation in (\ref{su2g1}) implies corresponding gauge transformations for $\Phi_{\sigma a}$:
\begin{align}
    \Phi_{\sigma a} \rightarrow \Phi_{\sigma b} U_{ba}^\dagger  \quad, \quad \Phi_{\sigma a} \rightarrow e^{-i \theta} \Phi_{\sigma a}\,. \label{su2g2}
\end{align}
The remaining Hamiltonian for $c_\sigma$, $F_a$, $\Phi_{\sigma a}$ is similar to the corresponding Hamiltonian for $c_\sigma$, $f_\sigma$, and $b$ in (\ref{eq:Hamiltonian}) for the Kondo lattice. So the main difference between the Kondo lattice case and the single-band case theories for FL* is that the U(1) gauge invariance of (\ref{eq:Hamiltonian}) has been replaced here by the SU(2)$\times$U(1) gauge invariances in (\ref{su2g1}) and (\ref{su2g2}). At the mean-field level, we assume an ansatz for the boson $\Phi_{\sigma a} \sim \Phi\, \delta_{\sigma a}$ \cite{ZhangSachdev_ancilla}, and then the ancilla theory is identical to the Kondo lattice theory in (\ref{eq:Hamiltonian}). 

In principle, we do have to also account for the coupling between the above Kondo lattice Hamiltonian for the top two layers, and the spin liquid layer. 
Within the FL* phase, this can be investigated perturbatively in $J_\perp$, as was examined in Ref.~\cite{Mascot22} (this paper also made connections to measured photoemission spectra in the cuprates). Alternatively, we can employ variational wavefunctions involving a projection onto rung singlets of the ${\bm S}_{1,2}$ layers. This was the approach employed in Refs.~\cite{Henry24,Iqbal24}, which also compared with observations on ultracold atomic systems. 

For the transition from FL* to FL, it was argued in Refs.~\cite{ZhangSachdev_ancilla,Zhang_Sachdev_ancilla2} that we need only account for the fluctuations of the $\Phi_{\sigma a}$ and associated SU(2)$\times$U(1) fluctuations. At the gaussian level, examined for the Kondo lattice in the present paper, there are no significant differences with the ancilla case. 

We can also consider confinement transitions out of the FL* into low temperature superconducting or charge-ordered phases. For these, it is essential to include the coupling to the bottom ${\bm S}_2$ layer. These transitions have been examined recently in Refs.~\cite{Christos1,Christos2,Christos3,BCS24,Zhang_Vortex24}.

\section{Material properties of Nd-LSCO}
\label{sec:dispersion}

Here we give details of the band structure of La$_{1.6-x}$Nd$_{0.4}$Sr$_x$CuO$_4$
as measured from experiments. The hopping parameters we used for Eq. (\ref{eq:dispersions c and b general}) are taken from Ref. \cite{Taillefer_Seebeck_PRX_2022} and are given by $t = 160 \, \text{meV}, ~ t' = -0.1364 \, t, ~ t'' = 0.0682 \, t, ~ \mu = -0.8243 \, t$.
Working in units of $t$ (i.e. setting $t = 1$), this gives the following values for the (spinless) density of states, transport function (for both spin species) and its derivative, all measured at the Fermi level: $    \nu_c = 0.3054, ~ \Phi^{(c)}_0 = 1.064, ~ \Phi'^{(c)}_0 = 0.83$. 

\section{Self-energies}
\label{sec:SE}

Here we discuss the computation of the self-energy for the conduction electrons. The calculation for the f-electrons is nearly identical. The entire self-energy is given by
\begin{align}
    \Sigma_c(i \o_n,T) &= g'^{2} T \sum_{i \o_m'} G_f(i \o_m' + i \o_n) G_b(-i\o_m')
    = - \frac{i}{2} \nu_f  g'^2 \frac{m_b}{2\pi} T \sum_{i \o_m'} 
    \sign(\o_m'+\o_n) \log(\frac{\L}{i \o_m' + \g \abs{\o_m'} + \Delta_b}),
\end{align}
where $\omega_n = 2 \pi T (n + 1/2),\omega_m' = 2 \pi T m$ are fermionic and bosonic Matsubara frequencies, respectively. We divide the Matsubara summation into parts, and, following some regrouping, the entire expression can be written as the sum of two terms 
\begin{equation}
\begin{split}
    & \Sigma_c(i \o_n,T)
    = - \frac{i}{2} \nu_f  g'^2 \frac{m_b}{2\pi} T
    \Bigg[
    \sum_{m = -\abs{n}+\Theta(-n-1/2)}^{\abs{n}-
    \Theta(-n-1/2)}
    \sign(2n + 2m +1) 
    \log(\frac{\L}{i 2 \pi m T + \g 2 \pi T   \abs{m} + \Delta_b})
    \\ & ~~~~~~~+
    \lim_{M\to\infty} \sum_{m = \abs{n}+\Theta(n+1/2)}^{M}
    \log(\frac{\L}{i 2 \pi m T + \g 2 \pi T   \abs{m} + \Delta_b})
    -
    \log(\frac{\L}{-i 2 \pi m T + \g 2 \pi T   \abs{m} + \Delta_b})
    \Bigg]
\end{split}
\end{equation}
Here $\Theta(x)$ is the Heaviside step function. Both summations can be performed in Mathematica. After rewriting in terms of $\omega_n$ and $z = \Delta_b/T$, the result is 
\begin{equation}
\begin{split}
   &  \Sigma_{c}(i\omega_n,T) = - i \gamma \frac{T m_b}{2 \nu_{c}} \Bigg[
    \sign{\omega_n} \left(
    \frac{\abs{\omega_n}}{\pi T} \log \left(\frac{\Lambda }{2 \pi  T \sqrt{1+\gamma^2}}\right)
    - \log \left(\frac{z}{2 \pi \sqrt{1+\gamma^2}}\right)
    - \log \abs{\frac{\Gamma \left(\f12 + \frac{z(\gamma+i)}{2 \pi (1+\gamma^2)} - \frac{\abs{\omega_n}}{2 \pi T} \right)}{\Gamma \left(\f12 + \frac{z(\gamma+i)}{2 \pi (1+\gamma^2)} \right)}}^2  
    \right)
    \\ & 
    + i \frac{\abs{\omega_n}}{\pi T} \arccot(\gamma)
    + \log \frac{\Gamma \left(\f12 + \frac{z(\gamma-i)}{2 \pi (1+\gamma^2)} + \frac{\abs{\omega_n}}{2 \pi T} \right)}{\Gamma \left(\f12 + \frac{z(\gamma+i)}{2 \pi (1+\gamma^2)} + \frac{\abs{\omega_n}}{2 \pi T} \right)}
    + \lim_{M\to\infty}
    \left\{
    i (2M+1) \arccot(\gamma)
    + \log \frac{\Gamma \left(M+1 + \frac{z(\gamma-i)}{2 \pi (1+\gamma^2)} \right)}{\Gamma \left(M + 1 + \frac{z(\gamma+i)}{2 \pi (1+\gamma^2)}\right)}\right\} 
    \Bigg],
\end{split}    
   \label{eq:Sigma_c,f full gamma not small}
\end{equation}
The first line (everything proportional to $\sign(\omega_n)$) comes from the first summation, while everything else is the result of the second summation. The last term inside of the curly brackets depends on the cutoff parameter $M$, and is independent of $\omega_n$. It can easily be checked that its contribution to $\Sigma_c$ is purely real, and therefore can be absorbed into the chemical potential. Analytically continuing to real frequency gives 
\begin{equation}
    \Sigma_{c,R}(x,T) = \gamma \frac{T m_b}{2 \nu_{c}} \Bigg[-\frac{x}{\pi} \log \left(\frac{\Lambda }{2 \pi  T \sqrt{1+\gamma^2}}\right) - i \frac{x}{\pi} \arccot(\gamma)  
    - i \log \left(\frac{\abs{\Gamma \left(\f12 + \frac{z(\gamma+i)}{2 \pi (1+\gamma^2)} \right)}^2}{\Gamma \left(\f12 + \frac{z(\gamma+i)}{2 \pi (1+\gamma^2)} - \frac{i x}{2 \pi} \right)^2} \frac{2\pi \sqrt{1+\gamma^2}}{z}\right)\Bigg] + C_{c}(T),
   \label{eq:Sigma_c,f retarded full gamma not small}
\end{equation}
where $x = \omega/T, z = \Delta_b/T$, and $C_{c}(T)$ is the frequency-independent real term discussed above. As we are only interested in the case when $\gamma \ll 1$ (c.f. Section \ref{sec: propagators}), this expression can be significantly simplified, which results in Eq. (\ref{eq:Sigma_c,f full}). Notably, neglecting these terms ($\gamma$ and $z(T) \gamma$) for the values of $\gamma$ we use in the main text is valid, as can be explicitly checked once the self-consistent expression for $z(T)$ (Eq. (\ref{eq:self-consistent boson gap equation})) is solved. We also note that taking the $T=0$ limit of this expression matches the same $T=0$ expression computed in Ref. \cite{Aldape2022}. 
\section{Vertex renormalizations}
\label{sec:vertices}

Here we discuss the renormalization of the two types of vertices: electrical and thermal. We show how the thermal vertex renormalizations cancel out. 

The $f$-electrons and boson share an emergent $U(1)$ charge, which is distributed between them. Once this emergent gauge field is integrated out, it leads to an intertwined renormalization of the current vertices for both species. Specifically, the renormalized currents are given by
\begin{align}
    &   
    \begin{pmatrix}
        J^{(f)}_R \\
        J^{(f)}_{Q,R} 
    \end{pmatrix}
    =
    \begin{pmatrix}
        - \Pi_b (\Pi_f + \Pi_b)^{-1} & 0\\
        - (\tilde \Pi_f -\tilde \Pi_b) (\Pi_f + \Pi_b)^{-1} & 1
    \end{pmatrix}
    \begin{pmatrix}
        J^{(f)} \\
        J^{(f)}_Q 
    \end{pmatrix}
    \\ &
    \begin{pmatrix}
        J^{(b)}_R \\
        J^{(b)}_{Q,R} 
    \end{pmatrix}
    =
    \begin{pmatrix}
        - \Pi_f (\Pi_f + \Pi_b)^{-1} & 0\\
        (\tilde \Pi_f -\tilde \Pi_b) (\Pi_f + \Pi_b)^{-1} & 1
    \end{pmatrix}
    \begin{pmatrix}
        J^{(b)} \\
        J^{(b)}_Q 
    \end{pmatrix}.    
\end{align}  
Here, the $\Pi_{\lambda}$ are electrical polarization bubbles, and $\tilde \Pi_{\lambda}$ are thermoelectric bubbles. 

The thermoelectric correlations function of the renormalized currents are then given by
\begin{align}
    \langle J^{(f)}_{Q,R} \, J^{(f)}_R \rangle
    & = - \(\tilde \Pi_f - (\tilde \Pi_f - \tilde \Pi_b)(\Pi_f + \Pi_b)^{-1} \Pi_f\) \, \Pi_b \, (\Pi_f + \Pi_b)^{-1}
    \\
    \langle J^{(b)}_{Q,R} \, J^{(b)}_R \rangle
    & = - \(\tilde \Pi_b + (\tilde \Pi_f - \tilde \Pi_b)(\Pi_f + \Pi_b)^{-1} \Pi_b\) \, \Pi_f \, (\Pi_f + \Pi_b)^{-1}    
\end{align}
Summing these up, the two large contributions (coming from the renormalization of $J^{(\lambda)}_{Q}$) cancel out, leading to
\begin{equation}
    \langle J^{(f)}_{Q,R} \, J^{(f)}_R \rangle
    + \langle J^{(b)}_{Q,R} \, J^{(b)}_R \rangle
    = - \(\tilde \Pi_f \, \Pi_b + \tilde \Pi_b \, \Pi_f\)\, (\Pi_f + \Pi_b)^{-1},
\end{equation}
which gives the expression for $L_1$ in Eq. (\ref{eq:L_0 and L_1}).

\twocolumngrid

\bibliography{references}

\end{document}